\documentclass{iau307}
\usepackage{graphicx}
\usepackage{natbib}
\usepackage{url}
\usepackage{dtklogos}
\usepackage{sidecap}
\bibpunct{(}{)}{;}{a}{}{,}

\title[Magnetospheres of Magnetic B-Type Stars] %% give here short title %%
{Plasma Leakage from the Centrifugal Magnetospheres of Magnetic B-Type Stars}

\author[Shultz et al.]   %% give here short author list %%
{Matt Shultz$^{1,2,3}$, Gregg Wade$^3$, Thomas Rivinius$^1$, \\ Jason Grunhut$^1$,  V\'eronique Petit$^4$ \\ \and the MiMeS Collaboration}

\affiliation{$^1$European Southern Observatory; $^2$Queen's University, Canada \\[\affilskip $^3$Royal Military College, Canada  ]
 $^4$University of Delaware, USA \\ email: {\tt mshultz@eso.org}}

\pubyear{2014}
\volume{307} 
\pagerange{}
\jname{New windows on massive stars: asteroseismology, interferometry, and spectropolarimetry}
\editors{G. Meynet, C. Georgy, J.H. Groh \& Ph. Stee, eds.}
\begin{document}

\maketitle

\begin{abstract}
Magnetic B-type stars are often host to {\em Centrifugal Magnetospheres} (CMs). Here we describe the results of a population study encompassing the full sample of known magnetic early B-type stars, focusing on those with detectable CMs. We present revised rotational and magnetic parameters for some stars, clarifying their positions on the rotation-confinement diagram, and find that plasma densities within their CMs is much lower than those predicted by centrifugal breakout. 
%% add here a maximum of 10 keywords, to be taken form the file <Keywords.txt>
\end{abstract}

\firstsection % if your document starts with a section,
              % remove some space above using this command.
\section{Scope}

Magnetic confinement of the winds of magnetic OB stars is explained by ud-Doula et al., (this volume) while for a description of CMs and the Rigidly Rotating Magnetosphere (RRM; \citealt{town2005a}) model we refer to Oksala et al. (this volume). 

It is not known how plasma escapes from CMs. The leading proposal is violent ejection in a so-called `centrifugal breakout' (CB) event \citep{town2005a, ud2008}: plasma density increases beyond the ability of the magnetic field to confine, thus rupturing the magnetic field structure. However, no direct evidence of CB has been found (e.g. \citealt{town2013}), motivating a deeper examination of the properties of CMs. 

\begin{figure}[t]
\begin{center}
%\centering
%\begin{tabular}{cc}
\includegraphics[width=1.\textwidth,natwidth=0.5]{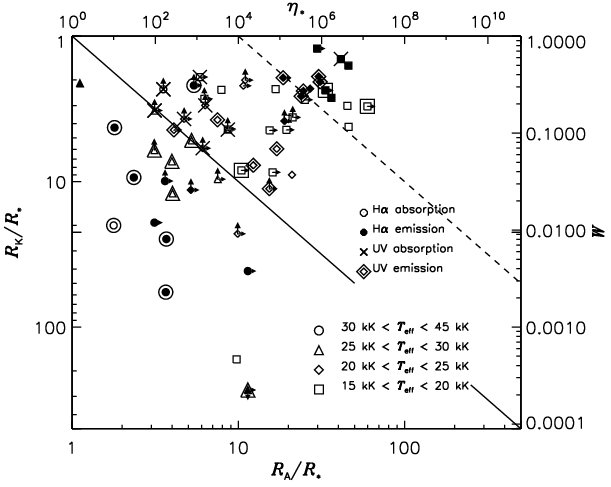} 
\caption{An up-dated version of the rotation-confinement diagram, first presented by \cite{petit2013}. In this revision, closer analysis of sample stars has moved some stars with H$\alpha$ emission above $R_{\rm A} = 10R_{\rm K}$ (the diagonal dashed line), while some stars with H$\alpha$ in absorption have been moved below this line. One H$\alpha$ absorption star, HD 35912, has been removed entirely, due to non-detection of a Zeeman signature in 11 ESPaDOnS observations with $<\sigma_B> \sim 40$ G. A new H$\alpha$ emission star, ALS 3694, has been discovered with new ESPaDOnS data. Finally, $R_{\rm K}$ of the early-type $\beta$ Cep star $\xi^1$ CMa has been revised sharply upward to $>$270 $R_*$ (see Shultz et al., this volume). }

%\end{tabular}

\label{fig1}
\end{center}
\end{figure}

We are conducting a population study of the magnetic B-type stars presented by \cite{petit2013} (see also Fig. 1), aimed at studying the dependence of CM emission on stellar, magnetic, and rotational properties. We are performing systematic follow-up observations of newly discovered or poorly studied magnetic B-type stars, with the intent of determining rotational periods (and hence Kepler radii $R_{\rm K}$) and dipolar magnetic field strengths (and hence Alfv\'en radii $R_{\rm A}$) for all stars. We also measure the emission strengths and plasma densities of the sub-set of stars with CMs via H$\alpha$ emission (about 25\% of the magnetic B-type population: filled symbols in Fig. \ref{fig1}).

%\begin{figure}[b]
%\begin{center}
%\includegraphics[width=1.\textwidth,natwidth=0.5]{Ha_UV_emew_mosaic.jpg} 
%\caption{ }
%\label{fig2}
%\end{center}
%\end{figure}

%\section{Emission Equivalent Widths\label{SecTwo}}
%

\section{Results\label{SecThree}}
We have measured the emission equivalent width (EW) of CM host stars, in both H$\alpha$ and UV resonance lines, by subtracting model photospheric spectra from observed spectra. We find clear thresholds dividing stars with and without H$\alpha$ emission ($R_{\rm A} > 20 R_*$, $R_{\rm K} < 3 R_*$, $\log{R_{\rm A}/R_{\rm K}} > 1$), however, beyond these thresholds there are no clear trends. In contrast to H$\alpha$, UV line emission is generally stronger, and more widely distributed in the rotation-confinement diagram (outlined symbols in Fig. \ref{fig1}). 

Using EWs for H$\alpha$, H$\beta$, and H$\gamma$, we measured the circumstellar plasma density via Balmer decrements \citep{ws1988} at emission maximum (corresponding to the rotational phase at which the line-of-sight is closest to perpendicular to the plane of the magnetosphere). There is no significant variation from the mean density, $\log{N}\sim 12.5$, with $0.15 < \sigma_{\log{N}} < 0.7$ dex. CB predicts both substantially higher plasma densities, and clear, strong trends in plasma density, increasing with increasing $R_{\rm A}$ and decreasing $R_{\rm K}$. Furthermore, CB predicts that plasma density should be high enough for H$\alpha$ emission in numerous stars for which no such emission is detected.

\bibliographystyle{iau307}
\bibliography{bib_dat.bib}

%\begin{discussion}

%\discuss{Author}{How do I know if I have something to put in the discussion section?}

%\discuss{Editors}{At the end of your talk, if somebody asked you a question, you will have to report it here.}

%\discuss{Authors}{How do I remember the questions?}

%\discuss{Editors}{During the conference, the people asking questions or comments were asked to write down their remarks. This paper should have been transmitted to you during the conference.} 

%\end{discussion}

\end{document}